\documentclass[12pt]{article}

\usepackage{graphicx}
\usepackage{dsfont}
\usepackage{amssymb}
\usepackage{amsmath}
\usepackage{mathrsfs}
\usepackage{bm}

\begin{document}

\section*{Comment on ``Quantum Friction --- Fact or Fiction?''}

\noindent
Ulf Leonhardt, University of St Andrews, North Haugh, St Andrews KY16 9SS, UK\\

\noindent
Quantum friction is neither fact nor fiction, but a theory. The theory \cite{Pendry,QF,VolPer} predicts that a dielectric moving relative to another dielectric experiences a force due to the quantum vacuum that is opposite to the direction of motion and would slow the dielectric down (Fig.\ 1A). There is no experimental evidence for or against this effect, no facts. Quantum friction is not fiction either, because a theory is based on and supported from other empirical evidence plus mathematics, extrapolated to a new subject of application. The question is whether in a theory all known relevant circumstances are taken into account. Pendry's paper \cite{Pendry} claims to establish an exact theory for a specific example of quantum friction. It qualitatively, but not quantitatively agrees with the previous literature \cite{QF,VolPer}, apart from papers \cite{PL} and \cite{Barton}. In our paper \cite{PL} we describe an exact calculation within the Lifshitz theory \cite{Lifshitz} of quantum forces and find exactly zero friction for general dielectrics.

Ultimately, only experiment is the judge, but in the absence of empirical facts, a few conclusions can still be drawn. First, in contrast to what is claimed in paper \cite{Pendry}, the theory developed there is not exact. Perturbation theory is applied and the paper ignores the polarization mixing by moving dielectrics. The latter is not an approximation, but an error, because polarization mixing already occurs in first order in the velocity, the same order as quantum friction. But there is a less technical argument why the argument of Pendry's paper \cite{Pendry} must be incorrect: quantum friction is impossible, because one could apply the same effect to extract an unlimited amount of useful energy from the quantum vacuum. The argument is briefly sketched in our previous paper \cite{PL}, but here it is explained in detail.

%%%
\begin{figure}[h]
\begin{center}
\includegraphics[width=30.0pc]{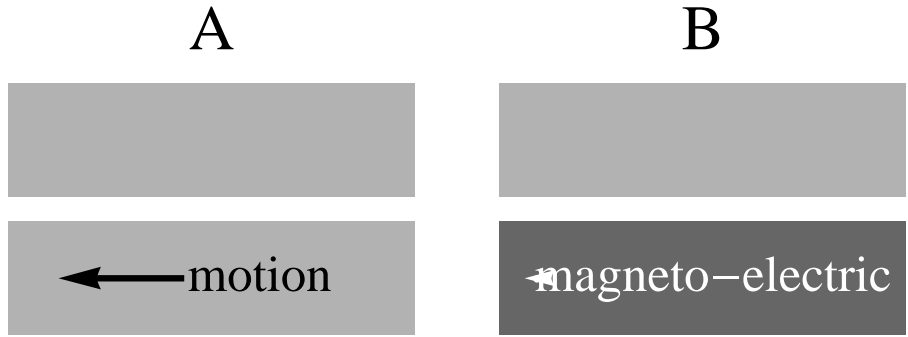}
\caption{
\small{
Quantum friction and magneto--electric acceleration. A: two dielectrics moving relative to each other are predicted \cite{Pendry,QF,VolPer} to experience a frictional force due to the quantum vacuum. B: as a moving medium is equivalent to a magneto--electric material, quantum friction should also exist between a magneto--electric and a piece of ordinary dielectric material such as glass. The glass would be accelerated by the quantum vacuum.
}
}
\end{center}
\end{figure}
%%%

\newpage
Consider, for simplicity, an impedance--matched dielectric with refractive index $n$ moving at velocity $\bm{u}$ relative to another dielectric. The moving medium obeys the constitutive equations \cite{LeoPhil}
%%%%%%
\begin{equation}
\bm{D} = \varepsilon_0\varepsilon\,\bm{E}
+ \frac{\bm{w}}{c}\times\bm{H}
\,,\quad
\bm{B} = \frac{\mu}{\varepsilon_0 c^2}\,\bm{H}
- \frac{\bm{w}}{c}\times\bm{E}
\label{c1}
\end{equation}
%%%%%%
with electric permittivity and magnetic permeability
%%%%%%%
\begin{equation}
\varepsilon = \mu = \frac{n}{1 - u^2 n^2 / c^2} \left( \left( 1 - \frac{u^2}{c^2} \right) \, \mathds{1}  +  \left( 1 - \frac{1}{n^2} \right) \frac{{\bm u} \otimes {\bm u}}{c^2}
\right) \approx  n \, \mathds{1}
\end{equation}
%%%%%%%
and magneto--electric coupling
%%%%%%%
\begin{equation}
{\bm w} = \frac{n^2 - 1}{1 - u^2 n^2 / c^2} \, {\bm u} \approx
(n^2 -1) \, {\bm u}  \, .
\label{c3}
\end{equation}
%%%%%%%
The constitutive equations (\ref{c1}-\ref{c3}) are derived from assuming the ordinary constitutive equations in the rest frame of the medium and Lorentz--transforming the fields. They show explicitly that electric and magnetic fields mix in moving dielectrics. The moving medium thus mixes the polarizations of light, the effect not considered in paper \cite{Pendry}. 

Moreover, Eqs.\ (\ref{c1}-\ref{c3}) are also the constitutive equations of a magneto--electric material \cite{Tretyakov} where the magneto--electric coupling plays the role of the velocity; here $\bm{u}$ is a parameter of the material. Hence, as far as electromagnetic fields are concerned, the moving medium is equivalent to a magneto--electric material. Quantum forces are generated by quantum fields, but as long as the constitutive equations are linear, they are also valid in quantum electrodynamics. Therefore, the quantum--electrodynamic effects of moving media are identical to the effects of a magneto--electric material with constitutive equations (\ref{c1}-\ref{c3}). The only possible difference between the two cases is the Doppler effect: the material of the moving medium responds to Doppler--shifted frequencies, whereas the magneto--electric medium experiences frequencies in the rest frame. But it turns out that the Doppler effect does not affect the following simple argument \cite{PL}.

Imagine a piece of an ordinary dielectric like glass brought close to a magneto--electric substrate with constitutive equations (\ref{c1}-\ref{c3}) (Fig.\ 1B). The substrate is assumed to be infinitely heavy, but not the glass. Relative to the glass, the magneto--electric is indistinguishable from a moving medium, as if the substrate were a conveyer belt for the glass. Quantum friction, if it existed, would act on both the glass and the apparently moving substrate, because motion is relative. It would cause a force on the substrate opposite to the direction of the parameter $\bm{u}$ and on the glass in the direction of $\bm{u}$. This quantum force would slow down the relative motion of the two bodies involved. As the substrate is heavy, but not the glass, the glass would be accelerated until its velocity matches the apparent velocity $\bm{u}$ of the substrate, as if the magneto-electric conveyer belt were dragging it along, but in reality the substrate is not moving at all. Still, setting the glass piece into motion would not be too surprising if it  had some source of internal energy like the spring of a toy car, but glass is a passive material. It also appears  \cite{Tretyakov} that the required magneto-electric material can be made, at least in principle, without an external field for $u<c/n$. Consequently, the theory \cite{QF} of quantum friction implies that the quantum vacuum could be a source of useful energy. One could put an arbitrary number of cleverly designed glass pieces on magneto-electrics and let them become accelerated by the quantum vacuum (Fig.\ 2). Quantum friction thus leads to the paradox that an unlimited amount of useful energy could be extracted from the quantum vacuum. 

%%%
\begin{figure}[h]
\begin{center}
\includegraphics[width=32.0pc]{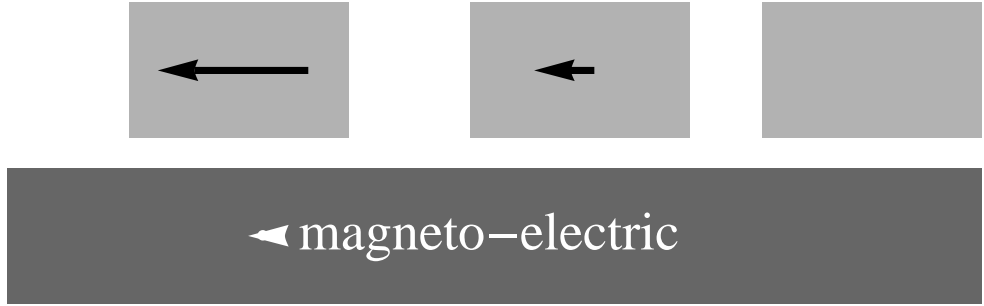}
\caption{
\small{
Perpetuum mobile. If quantum friction existed, an unlimited amount of useful energy could be extracted from the quantum vacuum by accelerating pieces of dielectric material above a magneto--electric substrate. 
}
}
\end{center}
\end{figure}
%%%

In our paper \cite{PL} Lifshitz theory \cite{Lifshitz} was applied to calculating the quantum forces between two dielectrics moving relative to each other. Lifshitz theory is a heavy theoretical machinery {\it a l\'a russe} far beyond the technical level of Pendry's paper \cite{Pendry}. Lifshitz theory states how the electromagnetic forces of the quantum vacuum can be expressed in terms of the electromagnetic Green function. It takes into account arbitrary electromagnetic properties of the material, as long as the material has a linear response, including dissipation. The main disadvantage of Lifshitz theory is that it is not very intuitive and computationally difficult, but if calculations can be performed Lifshitz theory is known to give reliable results that actually agree with experimental data where other, simpler methods fail. To put it simply, for calculating quantum forces, Lifshitz theory is the best we have got. 

There is no direct experimental evidence supporting the validity of Lifshitz theory for moving dielectrics, because the effects of motion are very small and precision measurements of quantum forces are incredibly difficult with present technology. But as Lifshitz theory assumes next to nothing about the material, apart from linear response, it should be extendable to moving media. In fact, a version of Lifshitz theory has been applied to calculations of quantum friction before \cite{VolPer}, but there the electromagnetic Green function was not correctly calculated; the papers \cite{VolPer} contain mathematical mistakes and partially contradict each other. In our paper \cite{PL} the Green function was exactly calculated. This Green function is valid for two arbitrary isotropic dielectric layers in uniform gliding motion relative to each other. Dispersion, dissipation in the materials and the electromagnetic effects of motion such as the Doppler effect and the mixing of electric and magnetic fields are fully taken into account. We found \cite{PL} that the diagonal part of the symmetrised Green function $G_s$ vanishes for all frequencies in the limit where the spectator point approaches the source point that should be taken in Lifshitz theory \cite{Lifshitz}. The frequency integral of $G_s$ gives the electric part of the Casimir stress \cite{Lifshitz}. It is convenient, for convergence issues, to move the integration contour to the imaginary half axis, but not necessary. The double curl of $G_s$ that gives the magnetic part  \cite{Lifshitz} is also diagonal in the appropriate limit \cite{PL}. The diagonality of the symmetrised GreenÕs function in this limit implies that there is no lateral force, no quantum friction \cite{PL}. Our paper \cite{PL} also makes a positive prediction about the modification of Casimir forces due to motion.  However, experimental tests of this theoretical prediction seem to be difficult within this century, because the effect is tiny, unless new technical developments arise.     

To summarize, in contrast to what is claimed, Pendry's paper \cite{Pendry} is not exact; it is based on perturbation theory and relevant physical effects such as polarization mixing are not taken into account. More importantly, if quantum friction existed an unlimited amount of useful energy could be extracted from the quantum vacuum and Lifshitz theory would fail. Both are unlikely to be true.


\begin{thebibliography}{99}

\bibitem{Pendry}
J.B. Pendry,
New J. Phys. {\bf 12}, 033028 (2010).

\bibitem{QF}
E.V. Teodorovich, Proc. Roy. Soc. A {\bf 362} (1978) and other papers cited in \cite{PL}. 

\bibitem{VolPer}
A.I. Volotichin and and B.N. Persson, 
J. Phys.: Condens. Matter {\bf 9}, 10301 (1999);
Phys. Rev. B {\bf 65}, 115419 (2002);
{\it ibid.} {\bf 74}, 205413 (2006);
{\it ibid.} {\bf 78}, 155437 (2008).

\bibitem{PL}
T.G. Philbin and U. Leonhardt, 
New J. Phys. {\bf 11}, 033035 (2009).

\bibitem{Barton}
G. Barton,
Ann. Phys. {\bf 245}, 361 (1996).

\bibitem{LeoPhil}
U. Leonhardt and T.G. Philbin, 
Prog. Opt. {\bf 53}, 69 (2009).

\bibitem{Lifshitz}
I.E. Dzyaloshinskii, E.M. Lifshitz, and L.P. Pitaevskii, Adv. Phys. {\bf 10}, 165 (1961); 
L.D. Landau, E.M. Lifshitz, and L.P. Pitaevskii, {\it Statistical Physics, Part 2} 
(Butterworth--Heinemann, Oxford, 1980).

\bibitem{Tretyakov}
A.N. Serdyukov, I.V. Semchenko, S.A. Tretyakov, and A. Sihvola, 
{\it Electromagnetics of bi--anisotropic materials: Theory and applications} 
(Gordon and Breach, Amsterdam, 2001).

\end{thebibliography}
\end{document}